\documentstyle[11pt]{article}

\newsavebox{\uuunit}
\sbox{\uuunit}
    {\setlength{\unitlength}{0.825em}
     \begin{picture}(0.6,0.7)
        \thinlines
        \put(0,0){\line(1,0){0.5}}
        \put(0.15,0){\line(0,1){0.7}}
        \put(0.35,0){\line(0,1){0.8}}
       \multiput(0.3,0.8)(-0.04,-0.02){12}{\rule{0.5pt}{0.5pt}}
     \end {picture}}
\newcommand{\dr}{\raise.3ex\hbox{$\stackrel{\leftarrow}{\delta }$}}
\newcommand{\dl}{\raise.3ex\hbox{$\stackrel{\rightarrow}{\delta}$}}
\newcommand{\beq}{\begin{equation}}
\newcommand{\eeq}{\end{equation}}
\newcommand{\ddr}{\raise.3ex\hbox{$\stackrel{\leftarrow}{d}$}}
\newcommand{\ddl}{\raise.3ex\hbox{$\stackrel{\rightarrow}{d}$}}

\def\gtwid{\raise.3ex\hbox{$>$\kern-.75em\lower1ex\hbox{$\sim$}}}
\def\ltwid{\raise.3ex\hbox{$<$\kern-.75em\lower1ex\hbox{$\sim$}}}

\newcommand{\frack}{\mbox{$\frac{1}{2}$}}
\newcommand{\fracq}{\mbox{$\frac{1}{4}$}}

\newcommand{\fabc}{\mbox{$f_{\alpha \beta \gamma}$}}

\newcommand{\nit}{\noindent}
\newcommand{\vs}{\vspace{2ex}}

\begin{document}

\pagestyle{plain}
\pagenumbering{arabic}

\begin{flushright}
DAMTP-95/37 \\
NIKHEF/95-049\\
\end{flushright}

\vspace{5ex}

\begin{center}
\Large{{\bf Particle Motion in a Yang-Mills Field:}} \\
\Large{{\bf Wong's Equations and Spin-$\frack$ Analogues$^*)$.}} \\
\vspace{5ex}

\large{{\bf N. \ Linden, A.J. \ Macfarlane }} \\
\vspace{3ex}

\large{ DAMTP, Cambridge University, Silver Street, Cambridge, U.K.} \\
\vspace{3ex}

\large {and} \\

\vspace{3ex}

\large{{\bf  J.W.\ van Holten}} \\
\vspace{3ex}

\large{ NIKHEF, Amsterdam NL} \\
\vspace{3ex}

July, 1995 \\
\vspace{5ex}

{\bf Abstract}

\end{center}
\vs

\begin{quote}\it

A complete, straightforward and natural Lagrangian
description is given for the classical non-relativistic
dynamics of a particle with colour or internal symmetry degrees of freedom
moving in a background Yang-Mills field. This provides a new simple
Lagrangian formalism
for Wong's equations for spinless particles, and presents also their
generalisation, in gauge covariant form,  for spin-$\frack$ particles,
within a complete Lagrangian formalism.

\end{quote}

\vspace{10ex}
\vspace{10ex}
\vspace{10ex}

\small{$^*)$ Presented by the second named author at the 4th Colloquium
"Quantum Groups and Integrable Systems", Prague, 22-24 June 1995.}

\newpage

\section{Introduction.}

\vs

Wong's equations are the equations of motion in the non-relativistic
mechanics of a spinless particle with colour or internal symmetry degrees of
freedom moving in a background Yang-Mills field.  The discussion originally
\cite{Wong} and subsequently has usually taken place
within a Hamiltonian formalism.  The
geometrical context of Wong's equations has been widely discussed
\cite{Ster} \cite{Wein} \cite{Mont} \cite{Mars}
using several versions of such a formalism.  But the long-standing problem
regarding the difficulties of displaying a satisfactory Lagrangian formalism
for the equations is still being
explicitly remarked upon \cite{Chan} at the present time.
Papers written about the problem, and its analogue for particles of
spin-$\frack$, divide roughly into two subsets,
according as whether or not Grassmann variables are used explicitly in
the description of spin or colour. Such variables do indeed
feature centrally in
\cite{BCL} \cite{BSSW} \cite{Arodz} \cite{SkagSt}
\cite{BMSS}, the last source being a monograph, which
reviews matters, provides collateral material and further references, as well
as discussing the problem in terms of bosonic variables. Other references
that also do so include
\cite{SkagSt} \cite{BBS}.

The present work actually differs in significant respects even from those of
the cited papers that do use
fermions to describe colour or spin. One main reason is
that it does not use either a
square root type of kinetic term in its starting Lagrangian, or else an
equivalent form that employs an einbein. It is in fact
non-relativistic from the outset, and any use that we make
of the Dirac formalism to take
account of constraints is trivial. Another aspect of our approach
that gives it distinct characteristics is its use of Majorana fermions to
describe spin and colour, using odd numbers of these where appropriate, and the
logically related use of $N=1$ supersymmetry with an odd number (one!) of
real Grassmann variables.
The latter is important because the treatment first of the supersymmetric
problem guides us towards suitable superactions, crucially
using spinorial
superfields to construct colour variables. This contrasts with the way that
the fermionic description of spin arises from the use of scalar superfields.
It also distinguishes our work from that of any previous paper that we have
found.

Valuable
background to the discussion of particle motion in a Yang-Mills field,
including explicit formulas for sourceless fields of this type,
and further references may be found in a recent interesting
monograph \cite{Carm}.

In this paper, we exhibit a
suitable Lagrangian for motion of a coloured particle in a Yang-Mills field
and deduce from it
the well-known Hamiltonian description
of Wong's equations.  Not only do we do this in the case of spinless
particles, but we also give the corresponding formalism
for particles of spin-$\frack$, presenting therein in gauge covariant form a
generalisation for spin-$\frack$ of Wong's equations.

We discuss the spin-$\frack$ case first, constructing a supersymmetric theory
with scalar and spinor superfields.  The former have the component expansions
\beq
\Phi_i=x_i+i\, \theta \phi_i \quad ,
\label{Phi}
\eeq
\nit
where the $x_i$ are the particle position co-ordinates (in three  spatial
dimensions).  Here $\theta$ is the single real Grassmann number of an $N=1$
supersymmetric theory which possesses a single real supercharge.
The $\phi_i$ are Majorana fermion variables.  The well known method
\cite{martin} \cite{Casal} \cite{BerMar}
\cite{DHV} \cite{dJMPvH}
of building a spin vector $S_i$ from them
\beq
S_i=-\frack i \varepsilon_{ijk} \phi_j \phi_k
\label{spin}
\eeq
\nit
guides us towards a suitable way of introducing colour variables. These come
from the Majorana fermion variables $\lambda_\alpha$ contained in the
spinor superfields
\beq
\Lambda_\alpha=\lambda_\alpha + \theta F_\alpha \quad ,
\label{Lamb}
\eeq
\nit
where $\alpha =1,2,\; \cdots \;,{\rm dim} \; g$, where
$g$ is a compact Lie algebra, the Lie algebra of the colour or internal
symmetry group of our theory.
The $F_\alpha$ feature as non-dynamical variables which we eliminate with the
aid of their Euler-Lagrange equations, leaving behind as dynamical variables
the $\lambda_\alpha$ out of which we construct colour variables or charges
$j_\alpha$ as in (\ref{spin}) by means of the definition
\beq
j_\alpha= -\frack  i \fabc\ \lambda_\beta \lambda_\gamma \quad .
\label{colo}
\eeq
The \fabc\ here are the totally anti-symmetric structure constants of the
Lie algebra $g$,
and $\lambda_\alpha$ and $j_\alpha$ transform according to its
adjoint representation.

It should be emphasised that, while the definition (\ref{colo}) is probably as
well known as the corresponding definition of spin, (\ref{spin}), our
use of spinorial superfields to bring the corresponding fermions,
$\lambda_\alpha$,
into the theory is clearly not.

We refer to a selection of papers \cite{saki} \cite{riet} \cite{macf}
where the general principles of constructing a supersymmetric theory with a
single Grassmann variable $\theta$  out of superfields (\ref{Phi}) and
(\ref{Lamb}) are described.  This leads naturally to a theory of a particle
with position $x_i$, spin $S_i$ and colour or charge $j_\alpha$ moving
in a background Yang-Mills field.  We find an action with supersymmetry
from which a complete canonical formalism emerges in a straightforward fashion.
This yields also a Hamiltonian formalism and Hamilton equations for $x_i$,
$S_i$ and $j_\alpha$ that are the generalisation of Wong's equations to the
case of particles of spin-$\frack$.
The fermion substructure of the spin and charge variables
is hidden in this Hamiltonian formalism and the equations of motion it gives.
But of course we now know it is nevertheless present and the key to the
associated Lagrangian formalism.  In fact, the supersymmetry is not essential
to the discovery of the Lagrangian formalism.  We have merely used it to
guide us towards a suitable type of Lagrangian.

It follows that we can pass to the treatment of the case of spinless particles
by omitting the $\phi_i$ variables. Considerations to do with supersymmetry
disappear at this point.  But we are lead directly to a Lagrangian and a
complete canonical formalism that contains the
previously well-understood Hamiltonian discussion of Wong's equations.
The Majorana fermions $\lambda_\alpha$ of the theory are again hidden
in the Hamiltonian formalism when this is finally couched in terms of the
Hamiltonian and $x_i$ and $j_\alpha$.  And again we now know that the fermion
substructure is lurking behind the scenes and essential to the Lagrangian
formalism.
\vs
\vs
\vs

\section{Supersymmetric Theory.}

\vs

We consider the theory governed by the action
\begin{eqnarray}
S & = & \int dt d\theta \, \frack \left( i \dot{\Phi_i} D\Phi_i \, +\,
\Lambda_\alpha D\Lambda_\alpha \, - \, g A_{i\alpha}(\Phi) \fabc \Lambda_\beta
\Lambda_\gamma D\Phi_i \right) , \nonumber \\
& = & \int dt d\theta (K+\theta L) \, = \, \int dt L \quad .
\label{Actn}
\end{eqnarray}
\nit
This involves the Lagrangian

\begin{eqnarray}
L & = & \frack \left( \dot{x_i} \dot{x_i}+i\phi_i \dot{\phi_i} + i
\lambda_\alpha \dot{\lambda_\alpha} +F_\alpha F_\alpha
+ig A_{i\alpha} \fabc \lambda_\beta  \lambda_\gamma \dot{x_i} \right)
\nonumber \\
  & + & \frack \left(
g\phi_j \phi_i A_{i\alpha ,j} \fabc \lambda_\beta \lambda_\gamma
+ig A_{i\alpha} \fabc (\lambda_\beta F_\gamma -F_\beta \lambda_\gamma) \right)
\quad .
\label{lagA}
\end{eqnarray}
The quantity $K$ seen in (\ref{Actn}) is used below in the construction of the
supercharge $Q$ by Noether's theorem.  The variables $F_\alpha$ do not appear
in our Lagrangian as dynamical.  Thus we can eliminate them using their
Euler-Lagrange equation.  This yields the result
\begin{eqnarray}
L & = & \frack \left( \dot{x_i} \dot{x_i} +i\phi_i \dot{\phi_i}
+i\lambda_\alpha
\dot{\lambda_\alpha} + igA_{i\alpha} \fabc \lambda_\beta \lambda_\gamma
\dot{x_i} \right) \nonumber \\
 & + & \fracq \,
F_{ij\alpha} \fabc \lambda_\beta \lambda_\gamma \phi_i \phi_j  ,
\label{lagB}
\end{eqnarray}
\nit
for $L$ in terms of dynamical variables.  Here
\beq
F_{ij\alpha}= \partial_i A_{j\alpha}- \partial_j  A_{i\alpha}
-g\fabc A_{i\beta}
A_{j\gamma} = \varepsilon_{ijk} B_{k\alpha} \quad ,
\label{curl}
\eeq
\nit
defines the usual covariant field variables.

The provision of the canonical formalism is problem free.  We find
\begin{eqnarray}
p_i & = & \dot{x_i}-gA_{i\alpha} j_\alpha \quad , \nonumber \\
\{ x_i, p_j \} & = & \delta_{ij} \quad , \,  \{ \lambda_\alpha, \phi_i \}=0
\quad , \nonumber \\
\{ \phi_i, \phi_j \} & = & -i\delta_{ij} \quad , \, \{\lambda_\alpha,
\lambda_\beta \}= -i\delta_{\alpha \beta} \quad ,
\label{PBs}
\end{eqnarray}
\nit
and the Hamiltonian
\beq
H=\frack (p_i+gA_{i\alpha} j_\alpha) \, (p_i +gA_{i\beta} j_\beta)
+g B_{k\alpha} j_\alpha S_k \quad ,
\label{Ham}
\eeq
\nit
where the notations (\ref{spin}) and (\ref{colo}), still to be justified,
have been used.  Using the well-known transformation properties of our
variables under supersymmetry transformations of Grassmann parameter
$\epsilon$, we apply Noether's theorem in the form
\beq
-i\epsilon Q=\sum \delta X \frac{\partial L}{\partial \dot{X}} -i\epsilon K
\quad ,
\label{noet}
\eeq
\nit
where the sum is over all the dynamical variables $X$.  This leads us
to the expression for the supercharge
\beq
Q=(p_i+gA_{i\alpha} j_\alpha) \phi_i \quad .
\label{supe}
\eeq
\nit
Using the consequence
\beq
\{ F,G \}= \frac{\partial F}{\partial x_k} \frac{\partial G}{\partial p_k}
-\frac{\partial F}{\partial p_k} \frac{\partial G}{\partial x_k}
+i(-)^f \frac{\partial F}{\partial \phi_k} \frac{\partial G}{\partial \phi_k}
+i(-)^f \frac{\partial F}{\partial \lambda_\alpha} \frac{\partial G}{\partial
\lambda_\alpha} \quad ,
\label{FG}
\eeq
where $f$ is the Grassmann parity of $F$,
\nit
to calculate Poisson brackets, we can show that $Q$ obeys the Poisson bracket
relation
\beq
\{ Q,Q \}=-2iH \quad .
\label{QQH}
\eeq
\nit
This is the classical analogue of the more familiar quantal
relationship $Q^2=H$.  We can also verify that $Q$ generates canonically the
supersymmetry transformation rules employed in the derivation of (\ref{supe}),
 a consistency check on the setting up of the formalism.
We can also use (\ref{FG}) to show that
\begin{eqnarray}
\{ S_i, \, S_j \} & = & \varepsilon_{ijk} S_k \quad , \nonumber \\
\{ j_\alpha, \, j_\beta \} & = & \fabc j_\gamma \quad ,
\label{Sj}
\end{eqnarray}
\nit
which justifies viewing $S_i$ and $j_\alpha$ as spin and colour or charge
variables.
In quantum theory in which we impose the
anticommutation relations $\phi_i \phi_j + \phi_j \phi_i= \delta_{ij}$,
we represent $\phi_i$ (in the
Schr\"{o}dinger) picture in terms of Pauli matrices by $\sigma_i/\sqrt{2}$.
It follows that we represent $S_i$ by $\sigma_i/2$, indicating that we are
dealing with the case of spin one-half.

We continue by using (\ref{FG}) to compute the time dependence of any classical
variable $Y$ via
\beq
\dot{Y}=\{ Y,H \} \quad .
\label{Hameq}
\eeq
\nit
We treat the variables $x_i,\dot{x_i},\phi_i,S_i,\lambda_\alpha$ and $j_\alpha$
in order.  The first calculation gives back the definition of $p_i$.
The others give rise to the equations of motion of the theory, the analogues
for  particles of spin one-half of Wong's equations, namely
\begin{eqnarray}
\ddot{x_i} & = & -g\varepsilon_{ijk} \dot{x_j} B_{k\alpha} j_\alpha
-gD_i B_{j\alpha} j_\alpha S_j \quad, \nonumber \\
\dot{S_k} & = & g j_\alpha B_{i\alpha} \varepsilon_{ijk} S_j \quad , \nonumber
\\
Dj_\alpha & = & g B_{i\beta} S_i \fabc j_\gamma \quad ,
\label{Eqsmot}
\end{eqnarray}
\nit
where
\begin{eqnarray}
Dj_\alpha & = & dj_{\alpha}/dt
+g \fabc \dot{x_i} A_{i\gamma} j_\beta \quad , \nonumber \\
D_i B_{j\alpha} & = & (\partial_i \delta_{\alpha \beta} +g \fabc A_{i\gamma})
B_{j\beta} \quad .
\label{coves}
\end{eqnarray}
\nit
The two terms on the right-hand side of the first equation of (\ref{Eqsmot})
have a simple physical interpretation. The first is a non-Abelian
generalisation of the Lorentz force and the second is a non-Abelian
Stern-Gerlach force, coupling the spin to the gradient of the $B$-field.

\nit

It makes perfect sense also to consider in its own right a theory using the
Hamiltonian specified in (\ref{Ham}) in terms of variables $x_i, S_i$
and $j_\alpha$, subject to the canonical relations $\{ x_i, p_j \} =
\delta_{ij}$, and (\ref{spin}) and (\ref{colo}). The same
equations (\ref{Eqsmot})
follow as the Hamilton equations of this dynamical theory too of course. The
fermionic substructure of spin and colour is hidden in this approach, and no
Lagrangian formalism suggests itself naturally.  This is indeed the
situation  that has obtained in the context of Wong's equations, i.e in the
spinless case,
for a long time.
\newpage

\section{Wong's Equations.}

\vs

There is no obstacle at all to adopting
the plan of simplifying the work of section two
by discarding the fermionic variables $\phi_i$ and ceasing to consider
anything to do with spin or supersymmetry.

Thus we shall set out in the theory of spinless particles from the Lagrangian
\beq
L_0= \frack \left( \dot{x_i} \dot{x_i} +i \lambda_\alpha
\dot{\lambda_\alpha} +ig A_{i\alpha} \fabc \lambda_\beta \lambda_\gamma
\dot{x_i} \right) \quad .
\label{lag0}
\eeq
\nit
We retain  the definition of $p_i$ contained in (\ref{PBs}), and employ the
canonical equations
\beq
\{ x_i,p_j \}=\delta_{ij} \quad , \{\lambda_\alpha , \lambda_\beta \}
=-i\delta_{\alpha \beta} \quad .
\label{PBs0}
\eeq
\nit
We retain also the definition (\ref{colo}) of $j_\alpha$.  Passage from $L_0$
to  the Hamiltonian $H_0$ gives rise to the expression
\beq
H_0=\frack (p_i+gA_{i\alpha} j_\alpha)(p_i+gA_{i\beta} j_\beta) \quad .
\label{Ham0}
\eeq
\nit
Calculating Hamilton's equations from (\ref{Ham0}) for the variables
$x_i,p_i,\lambda_\alpha$ and $j_\alpha$ in turn, we get the equations of motion
\begin{eqnarray}
\ddot{x_i} & = & -g\dot{x_j} F_{ij\alpha} j_\alpha \quad , \nonumber \\
Dj_\alpha & = & 0 \quad ,
\label{wongseqs}
\end{eqnarray}
\nit
where $F_{ij\alpha}$ and $Dj_\alpha$ are as defined before in (\ref{curl}) and
(\ref{coves}).  The equations (\ref{wongseqs})
agree exactly with Wong's equations, but here we have
derived them within a complete
canonical description of the classical mechanics of a spinless particle with
colour or charge vector $j_\alpha$ moving in a Yang-Mills field.  It is
complete in the sense that it stems in standard fashion from
a suitable starting Lagrangian, (\ref{lag0}).  It makes a good check on the
consistency of our work to verify that the equations of motion can also
be derived as the Lagrange equations of (\ref{lag0}).

We can also see easily that the historical description of Wong's equations is
contained in our work.  View $H_0$ as a function of the (Grassmann even)
variables $x_i,p_i$ and $j_\alpha$ subject only to the canonical equations
\beq
\{ x_i, p_j \}=\delta_{ij} \quad, \{j_\alpha, j_\beta \}=\fabc j_\gamma
\quad .
\label{redPBs}
\eeq
\nit
Hence, using
\beq
\{ F,G \}=\frac{\partial F}{\partial x_k} \frac{\partial G}{\partial p_k}
-\frac{\partial F}{\partial p_k} \frac{\partial G}{\partial x_k}
+ \frac{\partial F}{\partial j_\alpha} \frac{\partial G}{\partial j_\beta}
\fabc j_\gamma \quad ,
\label{lastPB}
\eeq
\nit
to compute Poisson brackets of (Grassmann even) functions of the dynamical
variables, we find, in  this way too, Wong's equations and very easily;
{\it e.g.}
\beq
dj_{\alpha}/dt=\{ j_\alpha, H_0 \} = \frac{\partial H_0}{\partial j_\beta}
\fabc j_\alpha =g\dot{x_i} A_{i\beta} \fabc j_\gamma \quad .
\label{jalpha}
\eeq
\nit
Although this last discussion is very simple indeed, it gives no clue to the
underlying fermionic nature of colour employed here to produce the sought after
Lagrangian formalism.


\newpage




\begin{thebibliography}{99}
\bibitem{Wong} S.K. \ Wong, Nuovo Cimento {\bf 65A} (1970) 689-694.
\bibitem{Ster} S. \ Sternberg, Proc. Nat. Acad. Sci. (U.S.A.) {\bf 74}
           (1977) 5253-5254.
\bibitem{Wein} A. \ Weinstein, Lett. Math. Phys. {\bf 2} (1978) 417-428.
\bibitem{Mont} R. \ Montgomery, Lett. Math. Phys. {\bf 8} (1984) 59-67.
\bibitem{Mars} J.E. \ Marsden and T.S. \  Ratiu, {\it Introduction to Mechanics
      and Symmetry}, Springer, Berlin, 1994.
\bibitem{Chan} Chan \  Hong-Mo, J. \ Faridani and Tsou Sheung Tsun,
      RAL Report, RAL-95-027, {\it A Non-Abelian Yang-Mills Analogue of
      Classical Electromagnetic Duality}.
\bibitem{BCL} A. \ Barducci, R. \ Casalbuoni and L. \ Lusanna, Nucl. Phys.
      {\bf B124} (1977) 93-108.
\bibitem{BSSW} A.P. \ Balachandran, P. \ Salomonson, B.-S. \ Skagerstam and
      J-O. \ Winnberg,  Phys. Rev {\bf D15} (1977) 2308-2317.
\bibitem{Arodz} H. \ Arod\'z, Acta Phys. Polonica {\bf B19} (1988) 697-708.
\bibitem{SkagSt} B.-S. \ Skagerstam and A. \ Stern, Physica Scripta {\bf 24}
      (1981) 493-497.
\bibitem{BMSS} A.P. \ Balachandran, G. \ Marmo, B.-S. \ Skagerstam and A. \
       Stern, {\it Gauge Symmetries and Fibre Bundles},  Lecture Notes in
       Physics {\bf 188}, Springer, Berlin, 1984.
\bibitem{BBS} A.P. \ Balachandran, S. \ Borchardt and A. \ Stern, Phys. Rev.
       {\bf D17} (1976) 3247-3256.
\bibitem{Carm} M. \ Carmeli, Kh. \ Huleihil and E. \ Leibowitz, {\it
             Gauge Fields, Classification and Equations of Motion},
             World Sci., Singapore, 1989.
\bibitem{martin} J.L. \ Martin, Proc. Roy. Soc. (London) {\bf 251} (1959)
          536-542 and 543-549.
\bibitem{Casal} R. \ Casalbuoni, Nuovo Cimento {\bf 33A} (1976) 389-431.
\bibitem{BerMar} F.A. \ Berezin and M.S. \ Marinov, Ann. Phys. {\bf 104} (1977)
          336-362.
\bibitem{DHV} E. \ D'Hoker and L. \ Vinet, Phys. Lett. {\bf 137B} (1984) 72-9.
\bibitem{dJMPvH} F. de Jonghe et al., NIKHEF-preprint, 1995.
\bibitem{saki} M. \ Sakamoto, Phys. Lett. {\bf 151B} (1985) 115-118.
\bibitem{riet} R.H. \ Rietdijk, PhD Thesis, Amsterdam (1992),
          {\it Applications of Supersymmetric Quantum Machanics}.
\bibitem{macf} A.J. \ Macfarlane, Nucl. Phys.\ {\bf B 438} (1995) 455-468.



\end{thebibliography}
\end{document}